\documentclass[cits]{PoS}\usepackage{amsmath,cite,color,bm,booktabs,dcolumn}

\newcolumntype{d}[1]{D{.}{.}{#1}}
\title{Decay Constants of the Heavy--Light Mesons $D^{(*)}$ and
$B^{(*)}$: Isospin Breaking}\ShortTitle{Decay Constants of the
Heavy--Light Mesons $D^{(*)}$ and $B^{(*)}$: Isospin Breaking}
\author{\speaker{Wolfgang Lucha}\\Institute for High Energy Physics,
Austrian Academy of Sciences, Nikolsdorfergasse 18, A-1050 Vienna,
Austria\\E-mail: \email{Wolfgang.Lucha@oeaw.ac.at}}
\author{Dmitri Melikhov\\Institute for High Energy Physics,
Austrian Academy of Sciences, Nikolsdorfergasse 18, A-1050 Vienna,
Austria, and\\ D.~V.~Skobeltsyn Institute of Nuclear Physics,
M.~V.~Lomonosov Moscow State University, 119991, Moscow, Russia,
and\\ Faculty of Physics, University of Vienna, Boltzmanngasse 5,
A-1090 Vienna, Austria\\E-mail: \email{dmitri\_melikhov@gmx.de}}
\author{Silvano Simula\\INFN, Sezione di Roma Tre, Via della Vasca
Navale 84, I-00146 Roma, Italy\\E-mail:
\email{simula@roma3.infn.it}} 

\abstract{QCD sum rules are analytic predictions of hadron
observables from quantum chromodynamics. In the so-called
local-duality limit of their Borel-transformed form, all
nonperturbative contributions are encompassed by just a single
quantity, an effective threshold beyond which, by assumption, all
hadronic contributions are reliably described by perturbative QCD.
Reconstructing its quark-mass dependence enables us to investigate
the differences of the leptonic decay constants of heavy--light
mesons effectuated by the tiny isospin-violating discrepancy of
the masses of up and down quarks. By and large, our findings enjoy
excellent agreement with the outcomes of earlier studies, the only
exception being a lattice-QCD prediction for the $B$ mesons, where
we find dramatic disagreement.}

\FullConference{The European Physical Society Conference on High
Energy Physics\\5--12 July 2017\\Venice, Italy}

\begin{document}QCD sum rules \cite{QSR} relate hadron properties
to the fundamental parameters of the quantum field theory of
strong interactions, quantum chromodynamics; they are found from
correlators of suitable interpolating operators by insertion of
complete sets of hadron states and use of the operator product
expansion. Applying Borel transformations removes all subtraction
terms and suppresses all hadron contributions above the ground
states. Any ignorance about heavier hadron states is swept under
the carpet by invoking quark--hadron duality which assumes QCD and
hadron contributions to mutually cancel above effective
thresholds. The accuracy \cite{LMSAU} of extracted predictions is
considerably raised if taking into account such thresholds'
dependence \cite{LMSET} on the parameters of the Borel
transformations. This advanced QCD sum-rule technique led us to
revisit the heavy--light-meson decay constants \cite{LMSDC},
specifically, their relative magnitude \cite{LMSR}. Recently, we
embarked on the analysis of isospin breaking induced in
heavy--light-meson decay constants by the tiny up and down quark
mass difference \cite{LMScf,LMSIB}.

In the local-duality limit of vanishing Borel variables, the QCD
sum rule for the decay constant $f_{H_q}$ of a meson $H_q$ formed
by a heavy quark $Q=c,b$ of mass $m_Q$ and a light quark $q=u,d,s$
of mass $m_q$ reduces to a dispersion integral of a spectral
density $\rho$ found perturbatively from QCD in form of a series
expansion in powers of the strong fine-structure constant
$\alpha_{\rm s}$, with all nonperturbative effects subsumed by its
effective threshold $s_{\rm eff};$ the inevitable truncation of
the perturbative series entails~an unphysical dependence of the
spectral density and the effective threshold on renormalization
scales:$$f_{H_q}^2=\int_{(m_Q+m_q)^2}^{s_{\rm eff}(m_Q,m_q)}{\rm
d}s\,\rho(s,m_Q,m_q,\alpha_{\rm s})\ .$$

Utilizing spectral densities up to $O(\alpha_{\rm s}\,m_q^1)$ and
$O(\alpha_{\rm s}^2\,m_q^0)$ \cite{CS,JL,G+}, we derive the decay
constants $f_P$ and $f_V$ of pseudoscalar mesons $P$ and vector
mesons $V,$ with masses $M_P$ and $M_V,$ defined in terms of the
axialvector and vector currents $A_\mu(x)=\bar
q(x)\,\gamma_\mu\,\gamma_5\,Q(x)$ and $V_\mu(x)=\bar
q(x)\,\gamma_\mu\,Q(x)$ according to
$\langle0|\,A_\mu(0)\,|P(p)\rangle={\rm i}\,f_{P}\,p_\mu$ and
$\langle0|\,V_\mu(0)\,|V(p)\rangle=f_{V}\,M_{V}\,\varepsilon_\mu(p).$
Using modified minimal subtraction renormalization for the
definition of parameters optimizes \cite{JL} perturbative
convergence, cf.~Table~\ref{Tab:n}.

\begin{table}[h]\begin{center}\caption{Numerical values of input
parameters in the modified minimal-subtraction renormalization
scheme.}\label{Tab:n}\vspace{2ex}\begin{tabular}{llr}\toprule
Quantity&Numerical input value&Reference\\\midrule$\alpha_{\rm
s}(M_Z)$&$0.1182\pm0.0012$&\cite{FLAG}\\$\frac{1}{2}\,(m_u+m_d)\equiv
m_{ud}(2\;\mbox{GeV})$&$(3.70\pm0.17)\;\mbox{MeV}$&\cite{FLAG}\\
$(m_d-m_u)(2\;\mbox{GeV})$&$(2.67\pm0.22)\;\mbox{MeV}$&\cite{FLAG}\\
$m_s(2\;\mbox{GeV})$&$(93.9\pm1.1)\;\mbox{MeV}$&\cite{FLAG}\\
$m_c(m_c)$&$(1.275 \pm 0.025)\;\mbox{GeV}$& \cite{PDG}\\
$m_b(m_b)$&$(4.247\pm0.034)\;\mbox{GeV}$&\cite{LMSmb}\\\bottomrule
\end{tabular}\end{center}\end{table}

\begin{figure}[t]\begin{center}
\includegraphics[scale=.3006]{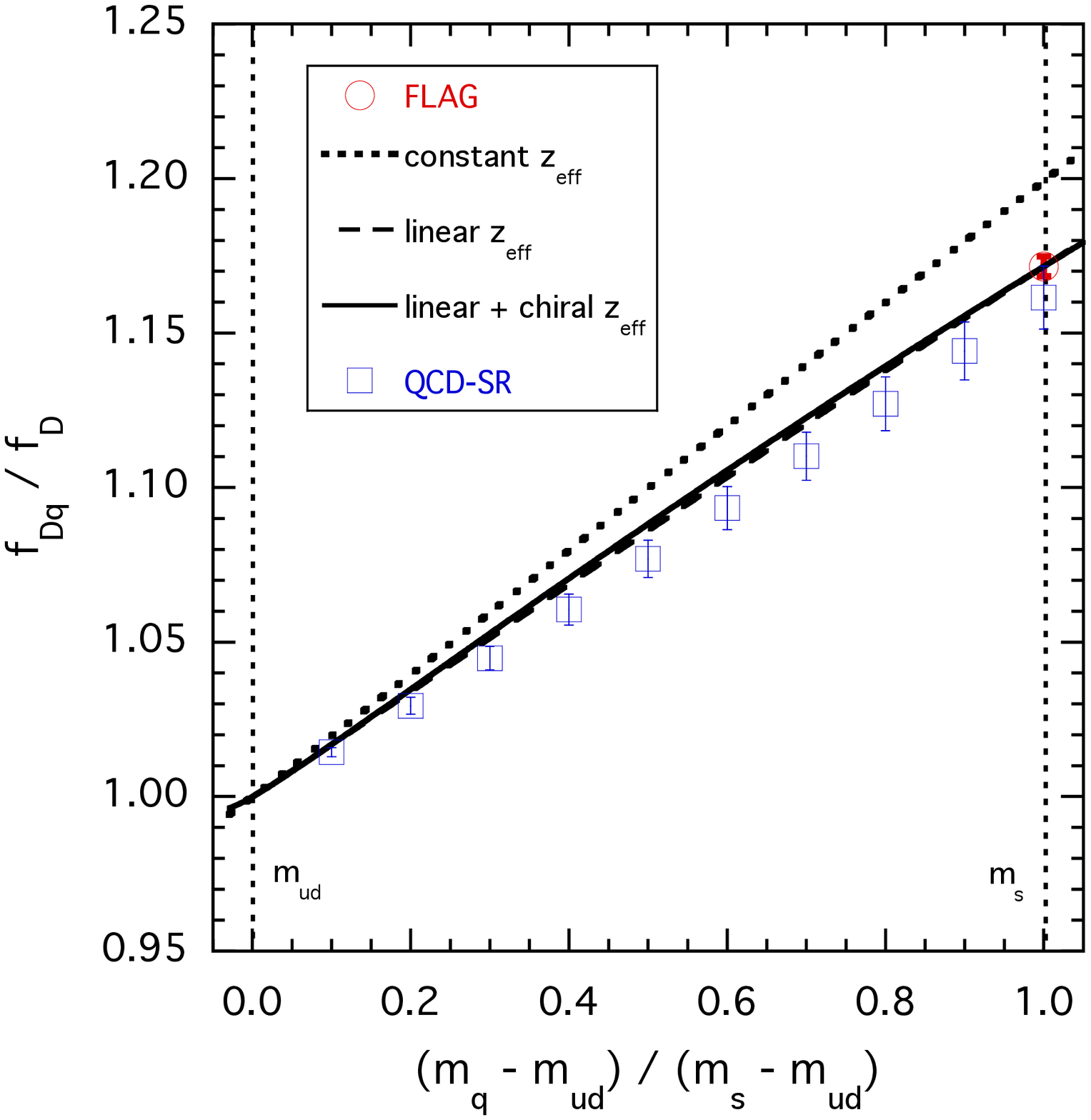}\qquad
\includegraphics[scale=.3006]{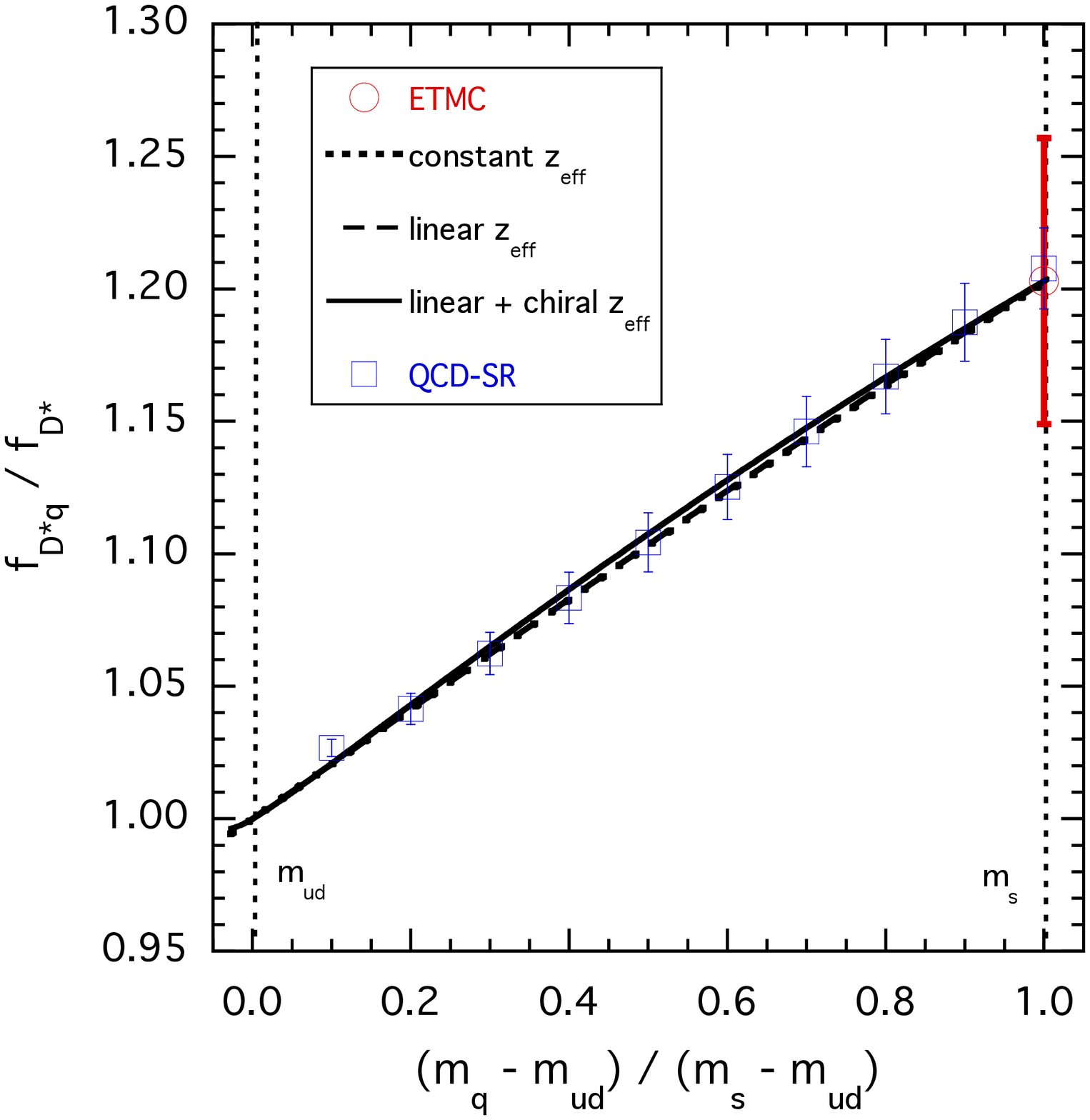}\\[1ex]
\includegraphics[scale=.3006]{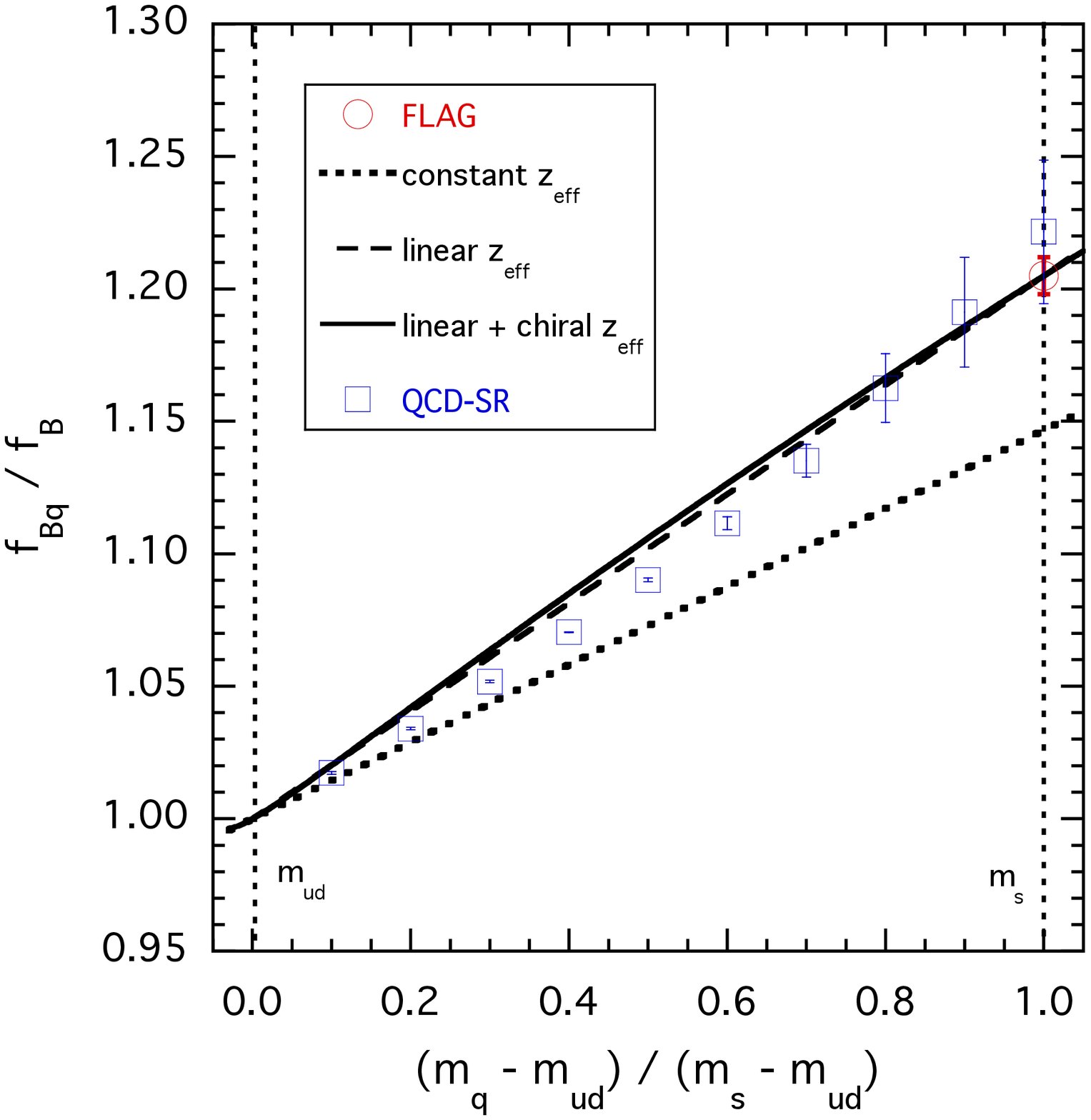}\qquad
\includegraphics[scale=.3006]{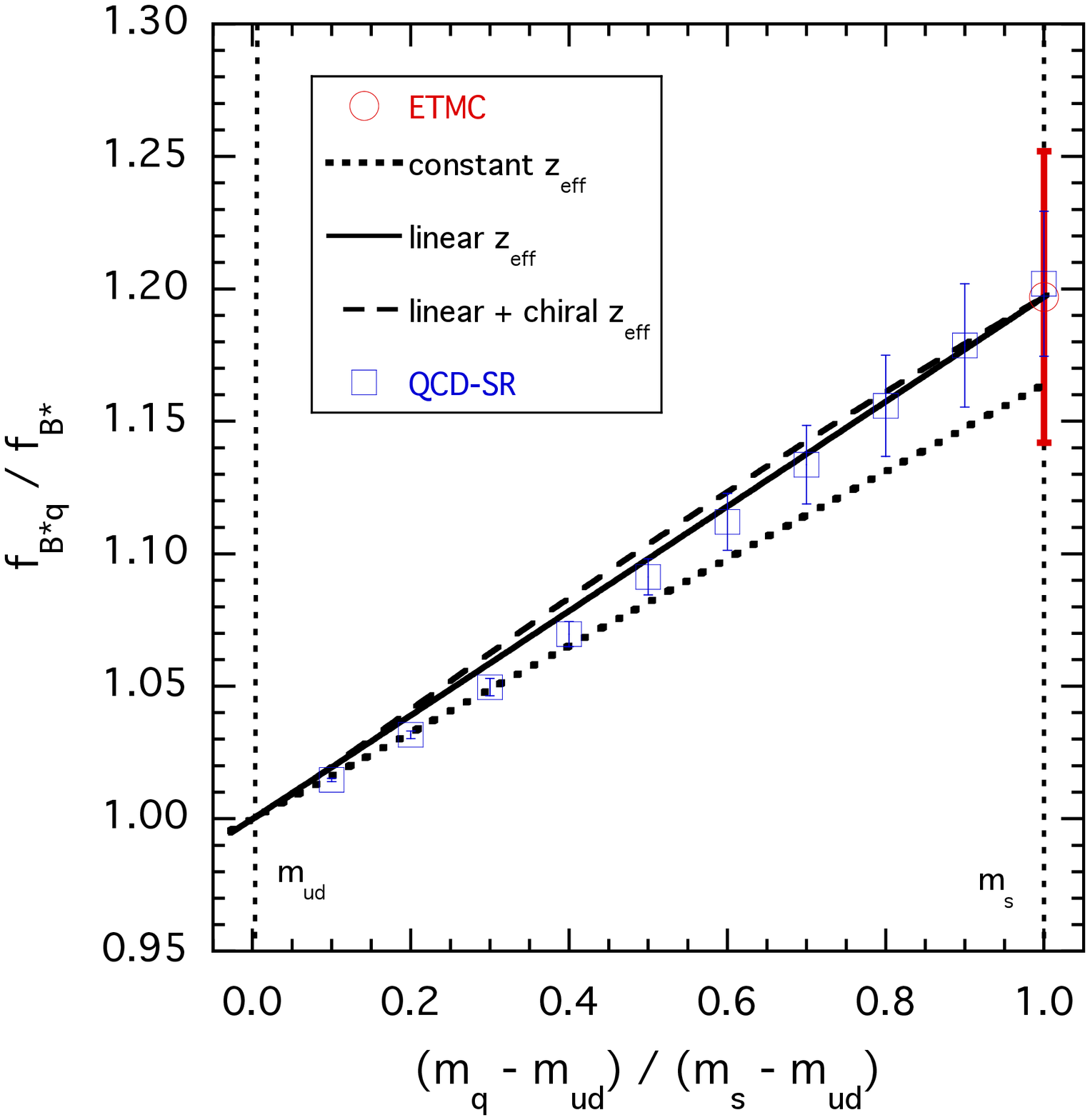}
\caption{Behaviour of $f_{H_q}/f_H(m_{ud})$ with redefined $m_q$
\cite{LMSLD}, compared with the results of Ref.~\cite{LMSIB}
(\textcolor{blue}{squares}).}\label{Fig:f}\end{center}\end{figure}

We fathom the discrepance $f_{H_d}-f_{H_u}$ of our heavy--light
meson decay constants by tracking the response $f_H(m_q)$ of the
local-duality QCD sum rule to continuous variations of the
light-quark mass $m_q$ in its interval $[0,m_s].$ Given the
perturbative spectral density at appropriately high order, the
only required ingredient is the effective threshold's quark-mass
dependence $s_{\rm eff}=s_{\rm eff}(m_Q,m_q).$ Shuffling together
available knowledge on heavy-quark expansion and chiral
logarithmic corrections \cite{SZ}, we distil parametrizations of
$z_{\rm eff}\equiv\sqrt{s_{\rm eff}}-m_Q-m_q$ \cite{LMSLD} the
coefficients of which are fixed by matching $f_H(m_q)$ for
$m_q=m_{ud}\equiv\frac{1}{2}\,(m_u+m_d)$ and $m_q=m_s$ to
lattice-QCD results \cite{FLAG,H*}. With respect to the $m_q$
dependence, we consider three ans\"atze for $z_{\rm eff}$:
constant, linear and linear plus chiral logarithms.\linebreak
Averaging over the two latter variants within the
renormalization-scale ranges $1$--$3\;\mbox{GeV}$ for charmed
mesons and $3$--$6\;\mbox{GeV}$ for bottom mesons converts the
resulting functional behaviour $f_H(m_q)$ (Fig.~\ref{Fig:f}) to
our prediction for the isospin-breaking differences of the
heavy-meson decay constants (Table~\ref{Tab:f}).

\begin{table}[b]\begin{center}\caption{Local-duality
isospin breaking results for heavy-meson decay constants,
exhibiting good agreement with Ref.~\cite{LMSIB} (apart from the
$D^{*\pm,0}$) and lattice-QCD outcomes for the $D^{\pm,0}$ but a
notable tension for the~$B^{0,\pm}.$}
\label{Tab:f}\vspace{2ex}\begin{tabular}{cd{1.8}d{1.8}d{1.12}}\toprule&
\multicolumn{3}{c}{$f_{H_d}-f_{H_u}\left[\mbox{MeV}\right]$}\\[.5ex]
\cline{2-4}&&\\[-2ex]Mesons $H_q$&\multicolumn{1}{c}{Borelized
threshold \cite{LMSIB}}& \multicolumn{1}{c}{Local-duality limit
\cite{LMSLD}}&\multicolumn{1}{c}{Lattice QCD}\\\midrule
$D^{\pm,0}$&0.97\pm0.13&0.96\pm0.09&0.94_{-0.12}^{+0.50}\hfill\mbox{\cite{LQD}}
\\$D^{*\pm,0}$&1.73\pm0.27&1.18\pm0.35&\multicolumn{1}{c}{---}\\
$B^{0,\pm}$&0.90\pm0.13&1.01\pm0.10&3.8\pm1.0\hfill\quad\mbox{\cite{LQB}}\\
$B^{*0,\pm}$&0.81\pm0.11&0.89\pm0.30&\multicolumn{1}{c}{---}\\\bottomrule
\end{tabular}\end{center}\end{table}

\noindent{\bf Acknowledgement.} D.M.~was supported by the Austrian
Science Fund (FWF), project no.~P29028.

\end{document}